\documentclass[prb,twocolumn,aps,showpacs,fixfloats]{revtex4}
\usepackage{graphicx}
\usepackage{bm}
\usepackage{amsmath,amssymb}
\usepackage{subfigure}
\usepackage{float}
\usepackage{latexsym}
\usepackage{color}
\usepackage{enumerate}
\usepackage{pdfpages}
\usepackage{tikz}
\usepackage{hyperref}
\usepackage{graphicx}
\usepackage{bm}
\usepackage{amssymb} 
\usepackage{amsmath}
\usepackage{subfigure}

\usepackage{relsize}

\begin{document}
\newcommand{\s}{\scriptscriptstyle}
\newcommand{\uu}{\uparrow \uparrow}
\newcommand{\ud}{\uparrow \downarrow}
\newcommand{\du}{\downarrow \uparrow}
\newcommand{\dd}{\downarrow \downarrow}
\newcommand{\ket}[1] { \left|{#1}\right> }
\newcommand{\bra}[1] { \left<{#1}\right| }
\newcommand{\bracket}[2] {\left< \left. {#1} \right| {#2} \right>}
\newcommand{\vc}[1] {\ensuremath {\bm {#1}}}
\newcommand{\tr}{\text{Tr}}
\newcommand{\Trans}{\ensuremath \Upsilon}
\newcommand{\Refl}{\ensuremath \mathcal{R}}


\title{High Landau levels  of  2D electrons  near the topological transition caused by interplay
of spin-orbit and Zeeman energy
 shifts }

\author{Rajesh K. Malla   and M. E. Raikh}

\affiliation{ Department of Physics and
Astronomy, University of Utah, Salt Lake City, UT 84112}

\begin{abstract}

In the presence of spin-orbit coupling two branches of the energy spectrum of 2D electrons get shifted in the
momentum space.  Application of in-plane magnetic field causes the splitting  of the branches in  energy.  When both, spin-orbit coupling  and 
Zeeman splitting are present, the branches of energy spectrum cross at certain energy. Near this energy,  the  Landau quantization becomes peculiar since  semiclassical trajectories, corresponding to individual  branches, get coupled.  We study this
coupling as a function of proximity to the 
topological transition. Remarkably, the dependence on the 
proximity is strongly asymmetric reflecting the specifics of the behavior of the trajectories near the crossing. Equally remarkable, 
on one side of the transition, the
magnitude of coupling is an oscillating function of this proximity. These oscillations can be interpreted in terms 
of the St{\"u}ckelberg interference.  Scaling of characteristic detuning with magnetic length is also unusual. This unusual behavior cannot be captured by simply linearizing the Fermi contours near the crossing point.


\end{abstract}

\maketitle

\section{Introduction}

It is known for more than 60 years  that, in a metal,  
the period of the resistance oscillations with magnetic  field 
as well as the period of the oscillations of  diamagnetic moment 
reflect the geometry of its Fermi surface.\cite{Onsager,Lifshitz1956} 
This relation originates from the fact that, by virtue  of the Landau quantization,    the areas of the 
cross-sections of the  Fermi surface by the planes perpendicular to magnetic field are discrete. These areas  are encircled in the course of  semiclassical motion of the electron wave packets in magnetic field and contain half-integer number of the flux quanta. 

In  particular situations when energy gaps, corresponding to
neighboring energy bands, are anomalously small, 
interband tunneling becomes possible.
This tunneling, known as magnetic breakdown,\cite{Zilberman,Azbel1961,Slutskin1968,KaganovSlutskin}    couples  the Fermi surfaces from different bands 
 and modifies the  quantization condition  to
\begin{equation}
\label{Quantization}
\cos \Bigg(\frac{S_{+}l^2+S_{-}l^2}{2} +\varphi_E \Bigg)
={\cal T}_E \cos\Bigg(\frac{S_{+}l^2-S_{-}l^2}{2}  \Bigg),
\end{equation}
where $S_{\pm}$ are the  areas encircled by the contacting semiclassical  trajectories, corresponding to the energy, $E$, and $l$ is the magnetic length. Parameters ${\cal T}_E$
and $\varphi_E$ are, respectively,
the amplitude and the phase of the coupling coefficient between the contacting trajectories.
The tunnel probability, 
$|{\cal T}_E|^2$, assumes an appreciable value at energies when the separation of the Fermi surfaces in the momentum space becomes  comparable to $l^{-1}$.  
Analytical form of ${\cal T}_E$ was established\cite{Zilberman} using the effective mass approximation,  within which the band dispersion near the touching  point has the form
\begin{equation}
\label{dispersion}
\varepsilon({\bf k})=\frac{\hbar^2k_x^2}{2m_x}- \frac{\hbar^2k_y^2}{2m_y},
\end{equation}
where $m_x$ and $m_y$ are the in-plane effective masses (magnetic field is directed along $z$).   As $E$ crosses from negative to positive  values, the connectivity of the Fermi surface, $\varepsilon({\bf k})=E$,  changes.  In magnetic field,  the tunneling 
probability between the states $k_x \rightarrow -\infty$ and  $k_x \rightarrow \infty$   reduces to the transmission through the  ``inverted parabola" potential,   $-\frac{\hbar^2}{2\left(m_xm_y\right)^{1/2}} \Big[\frac{(x-k_yl^2)}{l^2} \Big]^2$,  the result for   
which,  obtained in a celebrated paper by Kemble, \cite{Kemble1935}  reads
\begin{equation}
\label{TE}
|{\cal T}_E|^2=\frac{1}{\exp\left(-\pi \mu_E\right)+1},
\end{equation}
where  the parameter $\mu_E$ is proportional to energy  and is given by $\mu_E=\frac{1}{\hbar^2}\left(m_xm_y \right)^{1/2}El^2$.

Quantization condition Eq. (\ref{Quantization}) 
describes  topological transitions for spinless electrons with scalar wave-functions.
An alternative scenario of this 
transition\cite{Beenakker,G,G1,G2} unfolds in type-II Weyl semimetals 
predicted recently\cite{Bernevig} and realized experimentally, 
for review see Refs. \onlinecite{Review1},~\onlinecite{Review2}.
In these materials, 
the contacting contours of the Fermi surface belong to electron and hole pockets, see e.g. 
Ref.~\onlinecite{LifshitzTransition}. The corresponding states are the eigenfunctions 
of the matrix Hamiltonian, the simplest version 
of which has the form\cite{Beenakker}
\begin{equation}
\label{matrix}
{\hat H}_W=ak_x\sigma_0+\sum_{i}v_ik_i\sigma_i,
\end{equation}
where 
$\sigma_0$ is a unit matrix and $\sigma_i$ are the Pauli matrices.

Two branches,
\begin{equation}
\label{branches} 
E_{\pm}({\bf k})=ak_x\pm \left[\sum_i v_ik_i^2 \right]^{1/2},
\end{equation}
of the spectrum 
defined by the Hamiltonian Eq. (\ref{matrix}) touch at the point 
${\bf k}=0$. The difference between
the spectra Eq. (\ref{dispersion}) 
and Eq. (\ref{branches}) manifests itself
in the expression for the transmission probability. 
For type-II Weyl semimetals
it takes the form\cite{Beenakker}

\begin{equation}
\label{TW}
|{\cal T}_W|^2=\exp\left(-2\pi\mu_W\right),
\end{equation}
where $\mu_{W}$ is proportional to the square of  minimum separation between the contours and to the
square of magnetic length.
The origin of the difference between Eqs.
(\ref{TE}) and (\ref{TW})
 is that the Hamiltonian Eq. (\ref{matrix})   allows the Klein tunneling between
 the electron and hole states. 
With linear dispersion Eq. (\ref{branches}), the calculation of the tunnel probability reduces to the Landau-Zener problem.
 
 In Ref. \onlinecite{G3} it was noted that the   topological transition in the geometry of two
Fermi contours can be realized
for purely two-dimensional electrons subject to in-plane magnetic field and in the presence of spin-orbit coupling. The origin of  crossing of the two branches of the spectrum 
is the interplay of the Zeeman and spin-orbit splittings.\cite{Glenn} 
The eigenfunctions corresponding to the two crossing branches are spinors. Then it was concluded in 
Ref.     \onlinecite{G3}  that 
the semiclassical Landau quantization is governed
by Eq. (\ref{Quantization}) with tunnel probability given by Eq.~(\ref{TW}), similarly to the
type-II Weyl semimetals.

In the present paper we study in detail the evolution of the 2D Fermi contours 
in the vicinity of the topological transition emerging in the 
presence of  Zeeman and spin-orbit  couplings.
We show that, linearizing of the spectrum in the very vicinity
of crossing  is insufficient to describe the transition probability.
Magnetic field dependence of ${\cal T}$ as well 
as its dependence on detuning, is governed by the curvature of
the Fermi contours.

\section{ Evolution of the Fermi contours near  the crossing  }
\begin{figure}
	\includegraphics[scale=0.28]{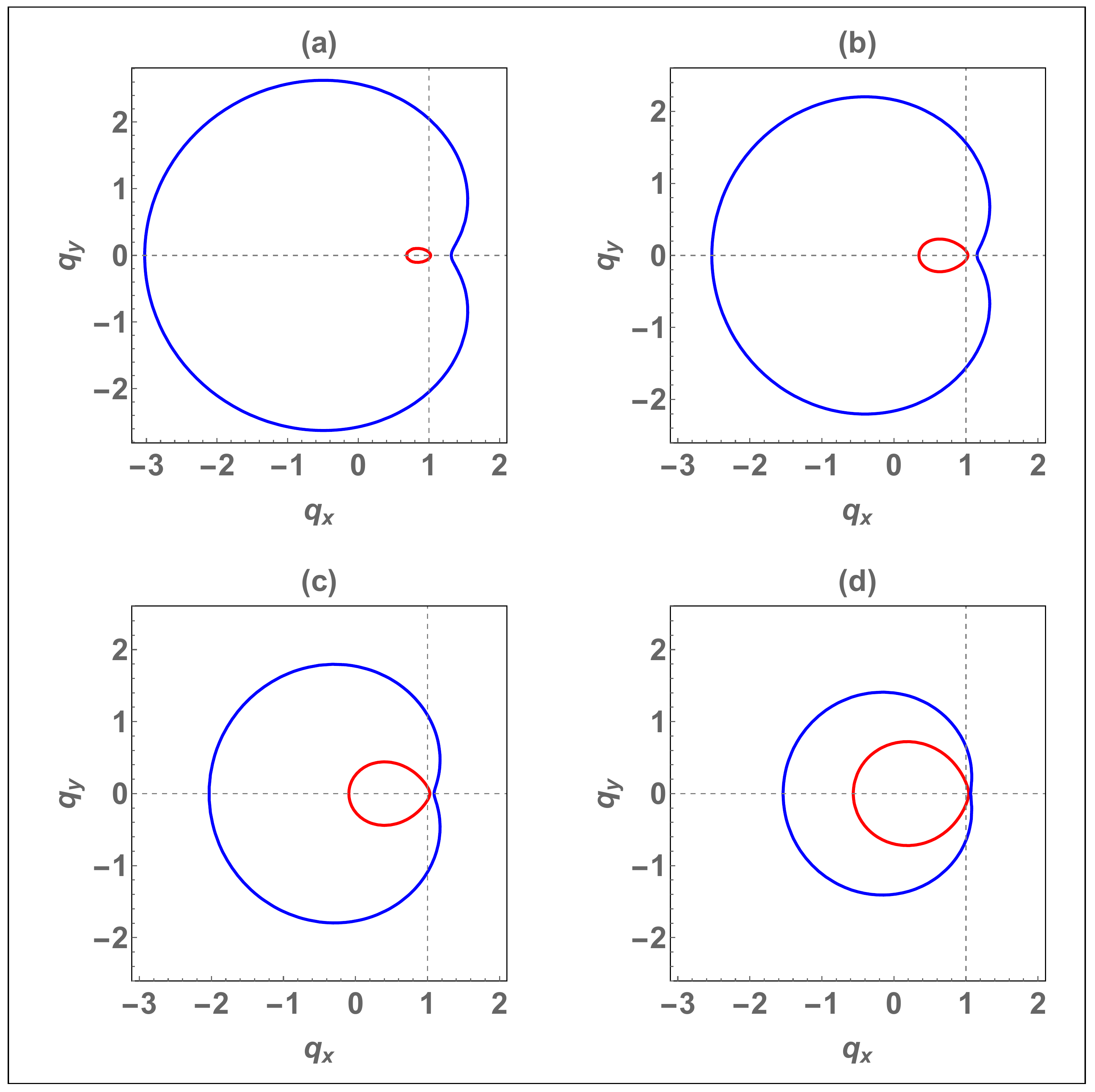}
	\caption{(Color online)  Evolution of the Fermi
	contours, defined by Eq. (\ref{dimensionless}), with the ratio, $\nu$, of the spin-orbit and Zeeman energy shifts [Eq. (\ref{nu})]; (a), (b), (c), and (d) correspond to $\nu=2$, $\nu=1.5$, $\nu=1$, and $\nu=0.5$, respectively. As $\nu$ decreases, the inner contour grows.   
	The Fermi energy is chosen to be ${\cal E}=1.1{\cal E}_0$ in all panels.  Thus, strictly speaking, the 
separation between the 	inner and the outer contours is finite at $q_y=0$. This separation can be distinguished in (a) and (b), but cannot be distinguished in (c) and (d).                                            }
\label{f1}
\end{figure}

\begin{figure}
	\includegraphics[scale=0.38]{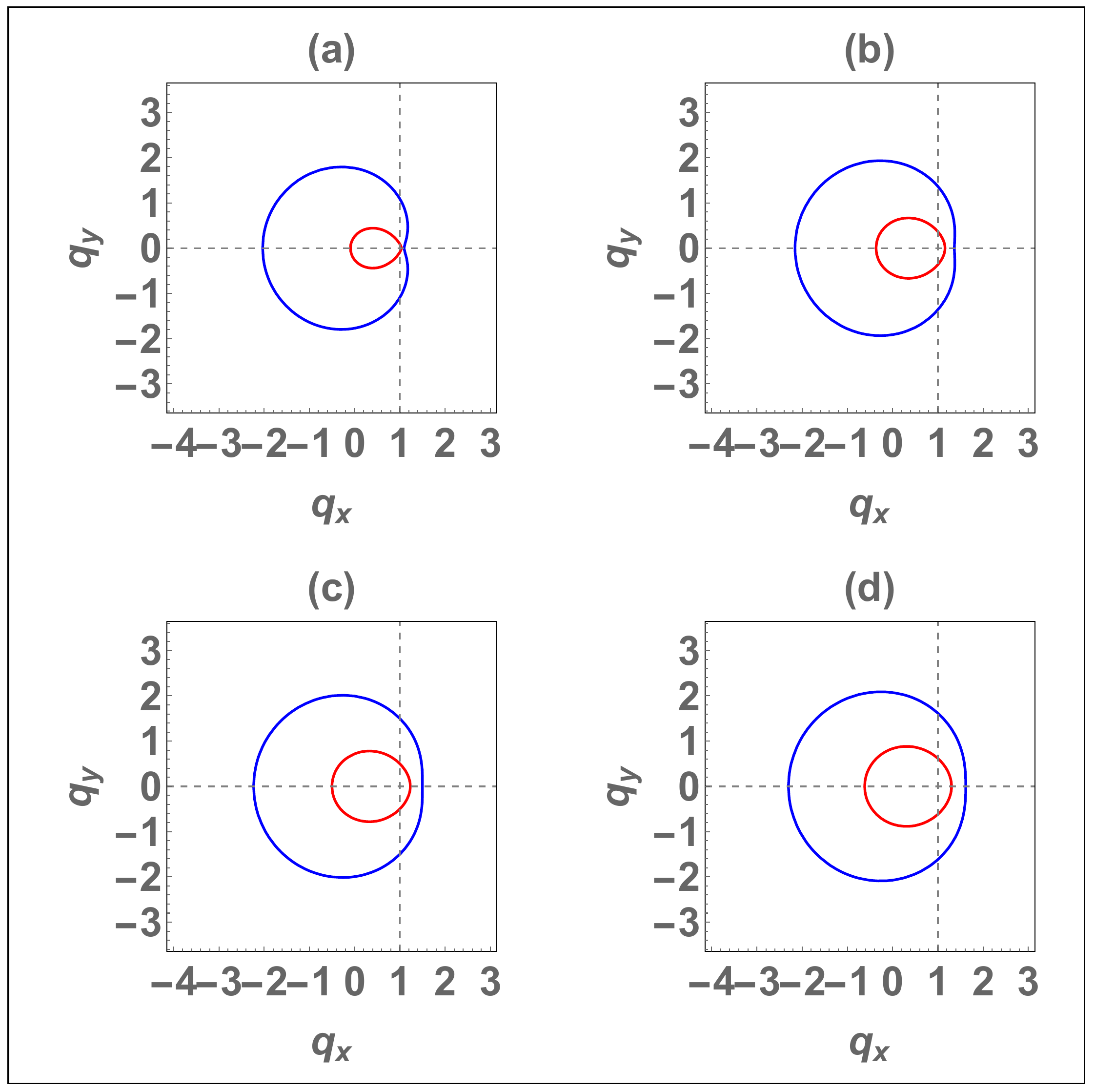}
	\caption{(Color online) Evolution of the Fermi
		contours, defined by Eq. (\ref{dimensionless}), with energy.   Panels (a), (b), (c), and (d)
		correspond to energies
		 ${\cal E}=1.1{\cal E}_0$ , ${\cal E}=1.5{\cal E}_0$, ${\cal E}=1.75{\cal E}_0$, and ${\cal E}=2{\cal E}_0$, respectively. We chose $\nu=1$ in all panels. It is seen that in panel (c) the outer contour 
	 is vertical at $q_y=0$ in accordance to Eq. (\ref{vertical}). }
	\label{f2}
\end{figure}
We start with a 2D Hamiltonian 
\begin{equation}
\label{H}
{\hat H}=\frac{\hbar^2 k^2}{2m}+
\alpha\left(k_x \sigma_y - k_y \sigma_x  \right)-\Delta\sigma_y,
\end{equation}
where the first term is a free-electron Hamiltonian, while the second and the third terms describe spin-orbit coupling and  Zeeman splitting in an in-plane magnetic field, respectively.  

Two branches of the spectrum of the Hamiltonian Eq.~(\ref{H}) are given by

\begin{equation}
\label{spectrum}
{\cal E}_{\pm}({\bf k})=\frac{\hbar^2(k_x^2+k_y^2)}{2m}
\pm \big[\left(\Delta-\alpha k_x\right)^2+\alpha^2k_y^2   \big]^{1/2}.
\end{equation}
The branches cross at  the point
\begin{equation}
k_x=\frac{\Delta}{\alpha}, ~~k_y=0,
\end{equation}
which corresponds to the energy 
\begin{equation}
{\cal E}={\cal E}_0=\frac{\hbar^2\Delta^2}{2m\alpha^2}.
\end{equation}
To analyze the behavior of the Fermi contours, ${\cal E}_{\pm}(\bf k)={\cal E}$, we introduce the
dimensionless variables 
\begin{equation}
k_x=\Bigg(\frac{\Delta}{\alpha }\Bigg)q_x,~~  k_y=\Bigg(\frac{\Delta}{\alpha}\Bigg)q_y,
\end{equation}
and rewrite Eq. (\ref{spectrum}) in the form

\begin{equation}
\label{dimensionless}
\frac{{\cal E}}{{\cal E}_0}=q_x^2+q_y^2 \pm \nu\big[ \left(q_x-1\right)^2+q_y^2       \big]^{1/2}, 
\end{equation}
where we have introduced a dimensionless parameter
\begin{equation}
\label{nu}
\nu=\frac{2m\alpha^2}{\hbar^2\Delta},
\end{equation}
which measures the ratio of the energy shifts due to the spin-orbit and Zeeman couplings.

Near the crossing point 
$\left({\cal E}-{\cal E}_0\right)\ll {\cal E}_0$ and
$q_y\ll 1$ Eq.~(\ref{dimensionless}) can be simplified to
\begin{equation}
\label{asymptotic}
\Bigg[q_x-1-\frac{2\left({\cal E}-{\cal E}_0\right)}{{\cal E}_0\left(4-\nu^2\right)}
\Bigg]^2\!-\!\Bigg(\frac{\nu^2}{4-\nu^2}   \Bigg)q_y^2
=\Bigg[\frac{\nu\left({\cal E}-{\cal E}_0  
\right)}{\left(4-\nu^2 \right){\cal E}_0}\Bigg]^2.
\end{equation}
We see that the behavior of the Fermi contours is
different for $\nu >2$ and for $\nu<2$. For $\nu>2$ Eq. (\ref{asymptotic})
describes an ellipse, i.e. there is only one Fermi contour.  
For $\nu<2$  two Fermi contours correspond to the  two branches of a hyperbola.  
 There  is a real crossing at 
${\cal E}={\cal E}_0$, namely, 
\begin{equation}
q_y=\pm \frac{\left( 4-\nu^2 \right )^{1/2}}{\nu}(q_x-1).
\end{equation}
Evolution of the Fermi contours with $\nu$ is illustrated in Fig. \ref{f1}. 
It is seen that,  as $\nu$ decreases  below $\nu=2$, the inner contours grows. 
The behavior of the outer contour is  quadratic near $q_y=0$ and 
also at two  finite values $\pm {\tilde q}_y$.
To find  these values,  we differentiate Eq. (\ref{dimensionless}) keeping
${\cal E}$ constant and obtain


\begin{equation}
\frac{\partial q_x}{\partial q_y}=-\frac{q_y\Bigg\{2\Big[\left(q_x-1\right)^2+q_y^2\Big]^{1/2} \pm \nu  \Bigg\}}
{ 2q_x\Big[ \left(q_x-1\right)^2+q_y^2 \Big]^{1/2} \pm \nu(q_x-1)                }.
\end{equation}
The sign ``$-$" corresponds to the outer branch. 
At $q={\tilde q}_y$ the derivative turns to zero, which, together with Eq.~(\ref{dimensionless}),
yields
\begin{equation}
\label{qx}
q_x={\tilde q}_x=\frac{1}{2}\Big(1+\frac{\nu^2}{4}+\frac{\cal E}{{\cal E}_0}   \Big),
\end{equation}
Substituting this value back into Eq.  (\ref{dimensionless}), we find

\begin{equation}
\label{vertical}
{\tilde q}_y
=\frac{1}{2}\Bigg[\Big(\frac{{\cal E}}{{\cal E}_0}+\frac{\nu^2}{4}+\nu-1\Big)
\Big(-\frac{{\cal E}}{{\cal E}_0}-\frac{\nu^2}{4}+\nu+1\Big)\Bigg]^{1/2}.
\end{equation}
We see that at energy $\frac{{\cal E}}{{\cal E}_0}=1+\nu-\frac{\nu^2}{4}$ 
the outer Fermi contour is vertical  at points 
(${\tilde q}_x$, $\pm {\tilde q}_y$), as illustrated in 
Fig.~ \ref{f2}. 
At small $\nu$ 
this energy is close to the crossing point of the
contours. In magnetic field, this peculiar behavior manifests itself
in the coupling between the semiclassical trajectories as we will see in the next Section.

\section{Tunneling between the semiclassical trajectories}
\begin{figure}
 	\includegraphics[scale=0.15]{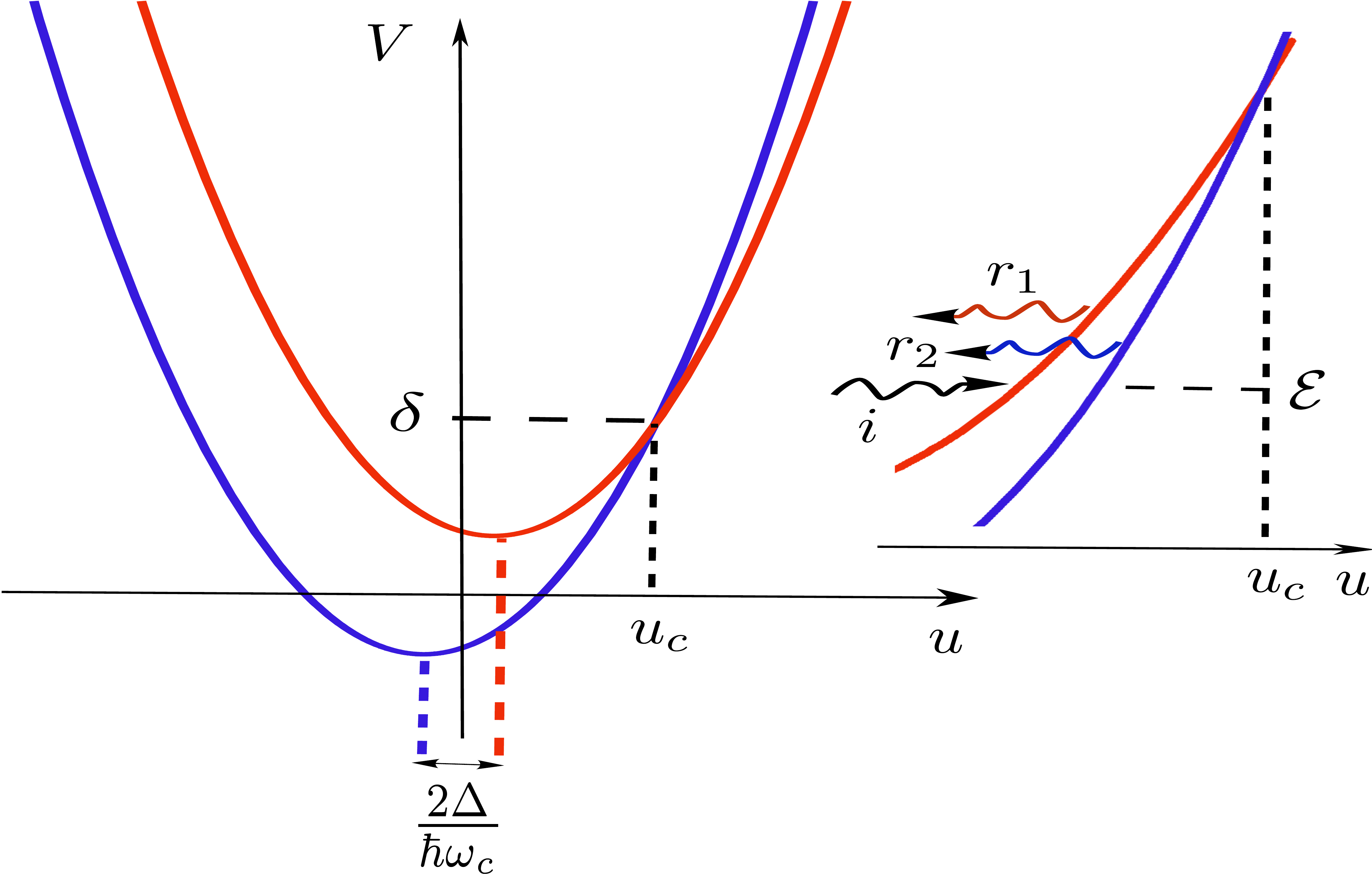}
 	\caption{(Color online)  Without the right-hand sides, Eqs. (\ref{Eq1}) and (\ref{Eq2}) are decoupled
 	and describe the electron motion in  parabolic potentials (blue and red, respectively).
 The potentials cross at $u=u_c$. Without coupling, the incident wave, $i$, see inset is fully reflected
into the wave, $r_1$, propagating in the red parabola. With right-hand sides caused by spin-orbit coupling,
another channel of reflection into the wave $r_2$ propagating in the blue parabola emerges. The corresponding reflection probability should be identified with transmission probability $|{\cal T}_E|^2$. }
 	\label{f4}
 \end{figure}

Incorporating  magnetic field in the $z$-direction amounts to  replacing 
$k_x$ by $k_x-\frac{y}{l^2}$. Then the system
of equations for the components of the spinor, $\left(\Psi_1,i\Psi_2\right)$,
takes the form
\begin{multline}
\label{first}
{\cal E}\Psi_1-\frac{\hbar^2}{2m}\left(k_x-\frac{y}{l^2}  \right)^2\Psi_1+\frac{\hbar^2}{2m}\frac{\partial^2\Psi_1}{\partial y^2}\\=
\alpha\left(k_x-\frac{y}{l^2}\right)\Psi_2-\Delta\Psi_2+\alpha\frac{\partial \Psi_2}{\partial y}, 
\end{multline}

\begin{multline}
\label{second}
{\cal E}\Psi_2-\frac{\hbar^2}{2m}\left(k_x-\frac{y}{l^2}  \right)^2\Psi_2+\frac{\hbar^2}{2m}\frac{\partial^2\Psi_2}{\partial y^2}\\=
\alpha\left(k_x-\frac{y}{l^2}\right)\Psi_1-\Delta \Psi_1-\alpha\frac{\partial \Psi_1}{\partial y}.
\end{multline}
Upon introducing new functions 
\begin{equation}
\Phi_1=\Psi_1+\Psi_2,~~~\Phi_2=\Psi_1-\Psi_2,
\end{equation}
and a dimensionless variable 

\begin{equation}
u=\frac{\Delta}{\alpha}\left(y-k_xl^2  \right)
\end{equation}
the system can be rewritten as
\begin{multline}
\label{Eq1}
\frac{{\cal E}}{{\cal E}_0}\Phi_1+\frac{\partial^2\Phi_1}{\partial
	u^2}-\Bigg[\Big(\frac{\alpha}{\Delta l}   \Big)^4\Big(u+\frac{\Delta}{\hbar \omega_c}   \Big)^2 -\Big(\frac{\nu^2}{4}+\nu  \Big)\Bigg]\Phi_1\\
=\nu \frac{\partial\Phi_2}{\partial u},
\end{multline}
\begin{multline}
\label{Eq2}
\frac{{\cal E}}{{\cal E}_0}\Phi_2+\frac{\partial^2\Phi_2}{\partial
	u^2}-\Bigg[\Big(\frac{\alpha}{\Delta l}   \Big)^4\Big(u-\frac{\Delta}{\hbar \omega_c}   \Big)^2 -\Big(\frac{\nu^2}{4}-\nu  \Big)\Bigg]\Phi_2\\
=-\nu \frac{\partial\Phi_1}{\partial u}.
\end{multline}
Here $\hbar\omega_c=\frac{\hbar^2}{ml^2}$ is the cyclotron energy. Equations  (\ref{Eq1}) and  (\ref{Eq2}) are obtained by adding and subtracting Eqs. (\ref{first}) and (\ref{second}).  Square brackets in 
Eqs. (\ref{Eq1}) and  (\ref{Eq2}) can be viewed as effective potentials for the functions $\Phi_1$ and $\Phi_2$. These potentials, sketched in Fig. \ref{f4} are parabolas shifted horizontally and vertically
These potentials cross at
\begin{equation}
u=u_c=\frac{\nu\hbar\omega_c}{2\Delta}\left(\frac{\Delta l}{\alpha}  \right)^4=\left(\frac{\Delta l}{\alpha}  \right)^2.
\end{equation}
The value of potential at $u=u_c$ is equal to

\begin{equation}
\delta=\frac{{\cal E}-{\cal E}_0}{{\cal E}_0}.
\end{equation}
Parameter $\delta$ is the dimensionless measure
of the proximity to the crossing. Semiclassical quantization
procedure is valid when the Landau levels, corresponding to
${\cal E}={\cal E}_0$, are high. Quantitatively, this condition
can be expressed as

\begin{equation}
\label{condition}
\frac{{\cal E}_0}{\hbar\omega_c}=\frac{1}{2}\Bigg( \frac{\Delta l}{\alpha}  \Bigg)^2\gg 1.
\end{equation}
If the above condition is satisfied,  derivation of the equation
similar to  Eq. (\ref{Quantization}) for the semiclassical energy levels can be outlined as follows. 
In the absence of the right-hand sides  in Eqs. (\ref{Eq1}), (\ref{Eq2}),  the solution of  (\ref{Eq1}) represents a wave,
incident from the left, which is fully reflected at the turning point (see Fig. \ref{f4}). The condition that
the solution decays to the right from the turning point defines the conventional  phase shift, $2\times \frac{\pi}{4}$,  between the incident
and reflected waves.   If the presence of the  right-hand side in Eq. (\ref{Eq2}),  
there are two channels of
reflection: in addition to the reflected-wave solution of Eq.  (\ref{Eq1}), the incident wave
can give rise to the solution of (\ref{Eq2}) propagating to the left, see Fig. \ref{f4}. If the amplitude
of the incident wave is $1$, then the amplitude of this second reflected wave should be identified with
${\cal T}_E$, the coupling coefficient in the quantization condition Eq.~ (\ref{Quantization}). 
Calculation of ${\cal T}_E$ is our main goal. 
To achieve this goal, it   is convenient to analyze the system Eqs. (\ref{Eq1}), (\ref{Eq2}) in the momentum space.

In the vicinity of $u=u_c$ the system (\ref{Eq1}), 
(\ref{Eq2}) takes the form
\begin{eqnarray}
\label{linearized}
\delta\Phi_1-{\cal F}_1u_1\Phi_1
+\frac{\partial^2\Phi_1}{\partial
	u_1^2}\!\!&=&\nu\frac{\partial \Phi_2}{\partial u_1},\nonumber
\\
\delta\Phi_2-{\cal F}_2u_1\Phi_2
+\frac{\partial^2\Phi_2}{\partial
	u_1^2}\!\!&=&\!\!-\nu\frac{\partial \Phi_1}{\partial u_1},\nonumber\\
\end{eqnarray}
where $u_1=u-u_c$. The slopes ${\cal F}_1$, ${\cal F}_2$
are defined as

\begin{equation}
\label{slopes}
{\cal F}_1=2\Big(\frac{\alpha}{\Delta l} \Big)^2
\Big(1+\frac{\nu}{2}\Big),~~~
{\cal F}_2=2\Big(\frac{\alpha}{\Delta l} \Big)^2
\Big(1-\frac{\nu}{2}  \Big).
\end{equation}
Upon performing the Fourier transformation in
Eq.~(\ref{linearized}), we
arrive to the system of coupled first-order 
differential equations for the transformed 
functions $\Phi_1$ and $\Phi_2$

\begin{eqnarray}
\label{system3}
\delta{\tilde\Phi}_1-i{\cal F}_1\frac{\partial{\tilde \Phi}_1}{\partial \kappa}-\kappa^2{\tilde\Phi}_1
&=&i\nu \kappa {\tilde \Phi}_2,\nonumber
\\
\delta{\tilde\Phi}_2-i{\cal F}_2\frac{\partial{\tilde \Phi}_2}{\partial \kappa}-\kappa^2{\tilde\Phi}_2
&=&-i\nu \kappa {\tilde \Phi}_1.
\end{eqnarray}
To analyze this system, it is 
convenient to ``antisymmetrize" it 
by eliminating the symmetric phase.
This is achieved by introducing instead of ${\tilde \Phi}_1$, ${\tilde \Phi}_2$ the new functions defined as
\begin{equation}
\label{symmetric}
{\tilde \Upsilon}_{1,2}(\kappa)={\tilde\Phi}_{1,2}(\kappa) \exp\Bigg[-i\left(\delta \kappa-\frac{\kappa^3}{3}\right)\frac{{\cal F}_1 +{\cal F}_2}{2{\cal F}_1{\cal F}_2} \Bigg].
\end{equation}
Then the system Eq. (\ref{system3}) assumes the form 
\begin{eqnarray}
\label{system4}
i{\cal F}_1\frac{\partial{\tilde \Upsilon}_1}{\partial \kappa}+\frac{{\cal F}_1 -{\cal F}_2}{2{\cal F}_2}\left(\delta-\kappa^2\right){\tilde \Upsilon}_1 =i\nu\kappa{\tilde \Upsilon}_2,\nonumber
\\
i{\cal F}_2\frac{\partial{\tilde \Upsilon}_2}{\partial \kappa}-\frac{{\cal F}_1 -{\cal F}_2}{2{\cal F}_1}\left(\delta-\kappa^2\right){\tilde \Upsilon}_2=-i\nu\kappa{\tilde \Upsilon}_1.
\end{eqnarray}
The product $\nu\kappa$ in the right-hand sides
describes the coupling between the semiclassical trajectories. We will first assume that the coupling is weak and find the transmission coefficient perturbatively. 
In the zeroth order we neglect the right-hand
side in the first equation, so that
\begin{equation}
{\tilde \Upsilon}_1(\kappa)=\exp\Bigg[i\frac{{\cal F}_1-{\cal F}_2}{2{\cal F}_1{\cal F}_2} 
\left(\delta \kappa -\frac{\kappa^3}{3}    \right)   \Bigg].
\end{equation}
Substituting ${\tilde \Upsilon}_1(\kappa)$ into the second equation and solving for ${\tilde \Upsilon}_2(\kappa)$ we find
\begin{equation}
\label{solution}
|{\tilde \Upsilon}_2(\infty)|^2=
\frac{\nu^2}{{\cal F}_2^2}
{\Bigg \vert}\int\limits_{-\infty}^{\infty}d\kappa~ \kappa
\exp\Big[i\frac{{\cal F}_1-{\cal F}_2}{{\cal F}_1{\cal F}_2} 
\left(\delta \kappa -\frac{\kappa^3}{3}       \right)\Big] ~ {\Bigg \vert}^2.
\end{equation}
The meaning of $|{\tilde \Upsilon}_2(\infty)|^2$
is the power transmission coefficient, 
$|{\cal T}_{{\cal E}}|^2$, analogously to Eq. (\ref{TE}).
\begin{figure}
	\includegraphics[scale=0.25]{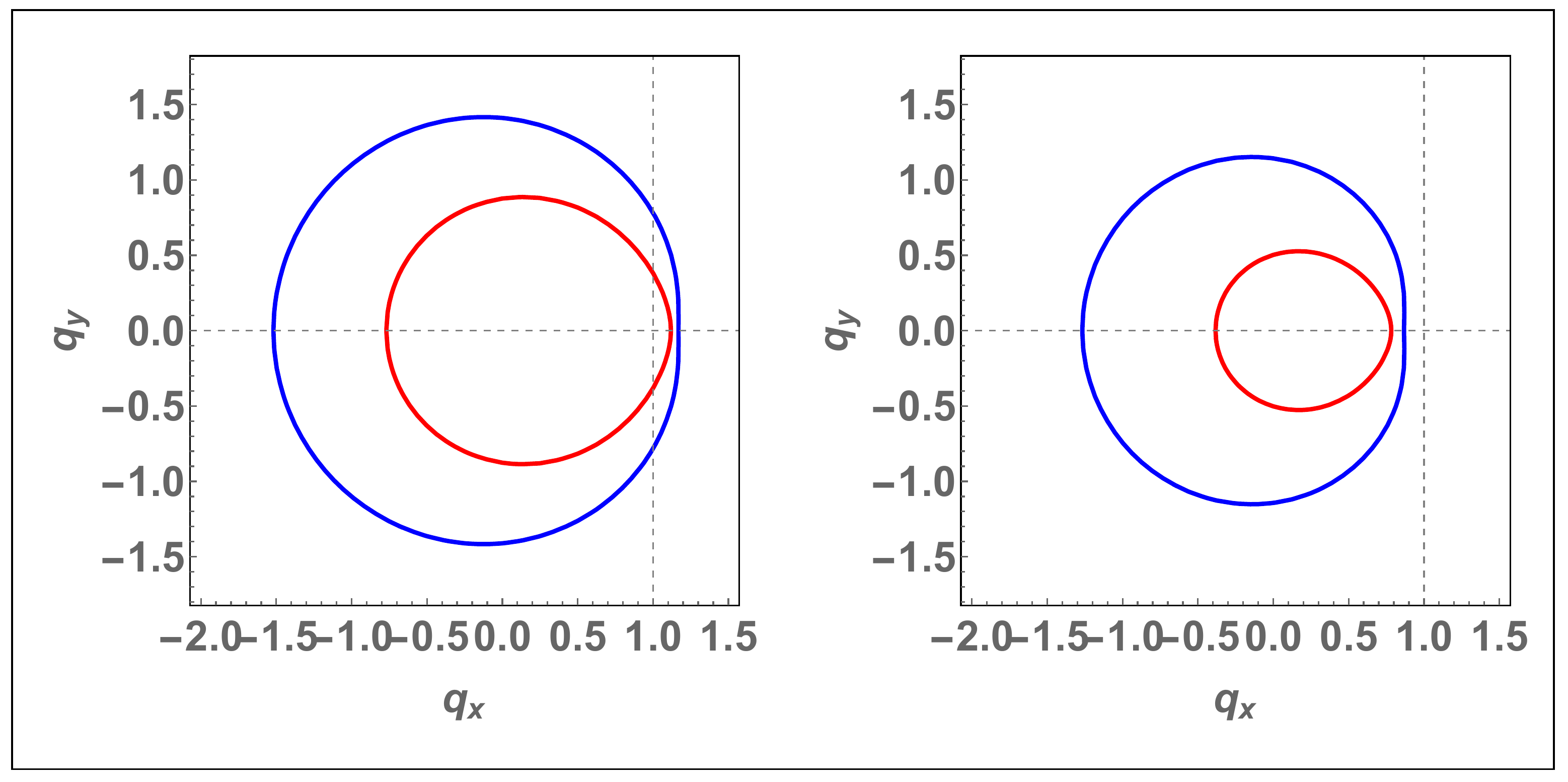}
	\caption{(Color online)  Illustration of the asymmetry of the Fermi-contours with respect to the 
		sign of detuning, $\delta$. In the left panel the contours are shown for $\delta=0.3$, while in the right
		panel for $\delta=-0.3$. Parameter $\nu$ is chosen to be $0.4$ for both panels. Despite the seemingly small
		difference between the two panels, the transmission coefficient, $|{\cal T}_E|^2$, in the right panel is much
		smaller than in the left panel.     }
	\label{f3}
\end{figure}

It is easy to see 
that only the imaginary part of the exponent contributes
to the integral. Then the integral reduces to
the derivative of the Airy function, $Ai(z)$. Using the expressions for ${
\cal F}_1$, ${\cal F}_2$ we rewrite the final result in the form 
\begin{equation}
\label{Airy}
|{\cal T}_{{\cal E}}|^2=4\pi^2\left(\frac{\nu}{2}\right)^{2/3}\left(\frac{\Delta  l}{\alpha}\right)^{4/3}
{\Bigg \vert} Ai'\left[-\delta \left(\frac{\nu}{2}\right)^{2/3}\left(\frac{\Delta  l}{\alpha}\right)^{4/3}\right]{\Bigg\vert}^2.
\end{equation}
Our prime observation is that the coupling is an asymmetric function of the detuning, $\delta$.
This, actually,  reflects the asymmetry of the Fermi-contours' arrangement with respect to the sign of $\delta$.
The situation is illustrated in Fig. \ref{f3}, where the Fermi contours are plotted for $\delta =0.3$
and $\delta =-0.3$. At negative  $\delta$ the transmission probability falls off with $|\delta|$
as $\exp\left[ -\frac{2}{3}|\delta|^{2/3} \nu\left(\frac{\Delta l }{\alpha}\right)^2\right]$. Note that, by contrast to Eqs. (\ref{TE}) and (\ref{TW}), characteristic $\delta$ scales with magnetic field as $l^{-4/3}$, instead of $l^{-2}$ and $l^{-1}$, respectively.

It is seen from Eq. (\ref{Airy}) that at positive $\delta$ the transmission coefficient oscillates with $\delta$. Unfortunately, Eq.~(\ref{Airy}) obtained perturbatively,  is not applicable in this domain. This is because it
predicts that $ |{\cal T}_{{\cal E}}|^2$ exceeds $1$ at large positive $\delta$. For this reason, in the next
Section we turn to numerics.

\section{Numerical results}
For numerical calculations it is convenient to perform a rescaling, $\kappa =Gz$, in the system Eq. (\ref{system4}),
where the parameter $G$ is equal to
\begin{equation}
\label{G}
G=\left( \frac{2}{\nu} \right)^{1/2}\left(\frac{\alpha}{\Delta l}\right)=\frac{\hbar\omega_c}{\Delta}.
\end{equation} 
Then the system assumes the form
\begin{eqnarray}
\label{first1}
i \tilde{\Upsilon}_1'+\frac{1}{2}\left(\frac{\delta}{G}-Gz^2\right)\tilde{\Upsilon}_1=iz\tilde{\Upsilon}_2, \nonumber\\
i \tilde{\Upsilon}_2'-\frac{1}{2}\left(\frac{\delta}{G}-Gz^2\right)\tilde{\Upsilon}_2=-iz\tilde{\Upsilon}_1.
\end{eqnarray}
We see that, effectively, the transmission coefficient depends only on  two parameters, detuning $\delta$ and the dimensionless magnetic field, $G^2$.
For numerical purposes it is convenient to get rid of the fast oscillations of ${\tilde\Upsilon_1}$ and ${\tilde \Upsilon_2}$ by
introducing new variables
\begin{equation}
\label{rho}
\rho_{1,2}=\tilde{\Upsilon}_{1,2}\exp\left[\mp \frac{i}{2} \left(\frac{\delta}{G}z-G\frac{z^3}{3}\right)\right].
\end{equation}
With these new variables the oscillating functions appear in the coupling of $\rho_1$ and $\rho_2$, namely
\begin{eqnarray}
\label{second1}
i\frac{\partial \rho_1 }{\partial z}=iz \exp\left[- i \left(\frac{\delta}{G}z-G\frac{z^3}{3}\right)\right]\rho_2, \nonumber\\
i\frac{\partial \rho_2 }{\partial z}=-iz \exp\left[ i \left(\frac{\delta}{G}z-G\frac{z^3}{3}\right)\right]\rho_1.
\end{eqnarray}
In terms of parameter $G$, the result Eq. (\ref {Airy}) reads
\begin{equation}
\label{Airy1}
|{\cal T}_{{\cal E}}|^2=|\rho_2(\infty)|^2=\frac{4\pi^2}{G^{4/3}}
{\Bigg \vert} Ai'\left(-\frac{\delta}{G^{4/3}}\right){\Bigg\vert}^2.
\end{equation}

In our numerical calculations we first analyzed the behavior of $|\rho_{1,2}|^2$ with $z$. In general these quantities exhibit oscillations
on the background of a smooth envelop. There is way to approximately isolate this envelop.   To do so,
we integrate the second equation of the system
Eq. (\ref{second}) using the condition 
$\rho_2  (-\infty)=0$ and substitute the expression for $\rho_2(z)$ into the first equation. This yields the following closed integral-differential equation for $\rho_1(z)$ 
 
\begin{equation}
\label{integral}
\frac{\partial \rho_1 }{\partial z}=z\int\limits_{-\infty}^z dz' z'\rho_1(z')  \exp\Bigg\{i \left[ \frac{\delta}{G} \left(z-z'\right) -\frac{G}{3}\left(z^3-z'^3\right)\right]\Bigg\}.
\end{equation}
The procedure of extracting the envelop from this equation is developed in   Ref. \onlinecite{Malla}. Employing  this procedure  yields
\begin{equation}
\label{differential}
|\rho_1(z)|^2=\frac{\left(\frac{\delta}{G}-Gz^2\right)^2}{z^2+ \left(\frac{\delta}{G}-Gz^2\right)^2}.
\end{equation}

\begin{figure}
		\includegraphics[scale=0.28]{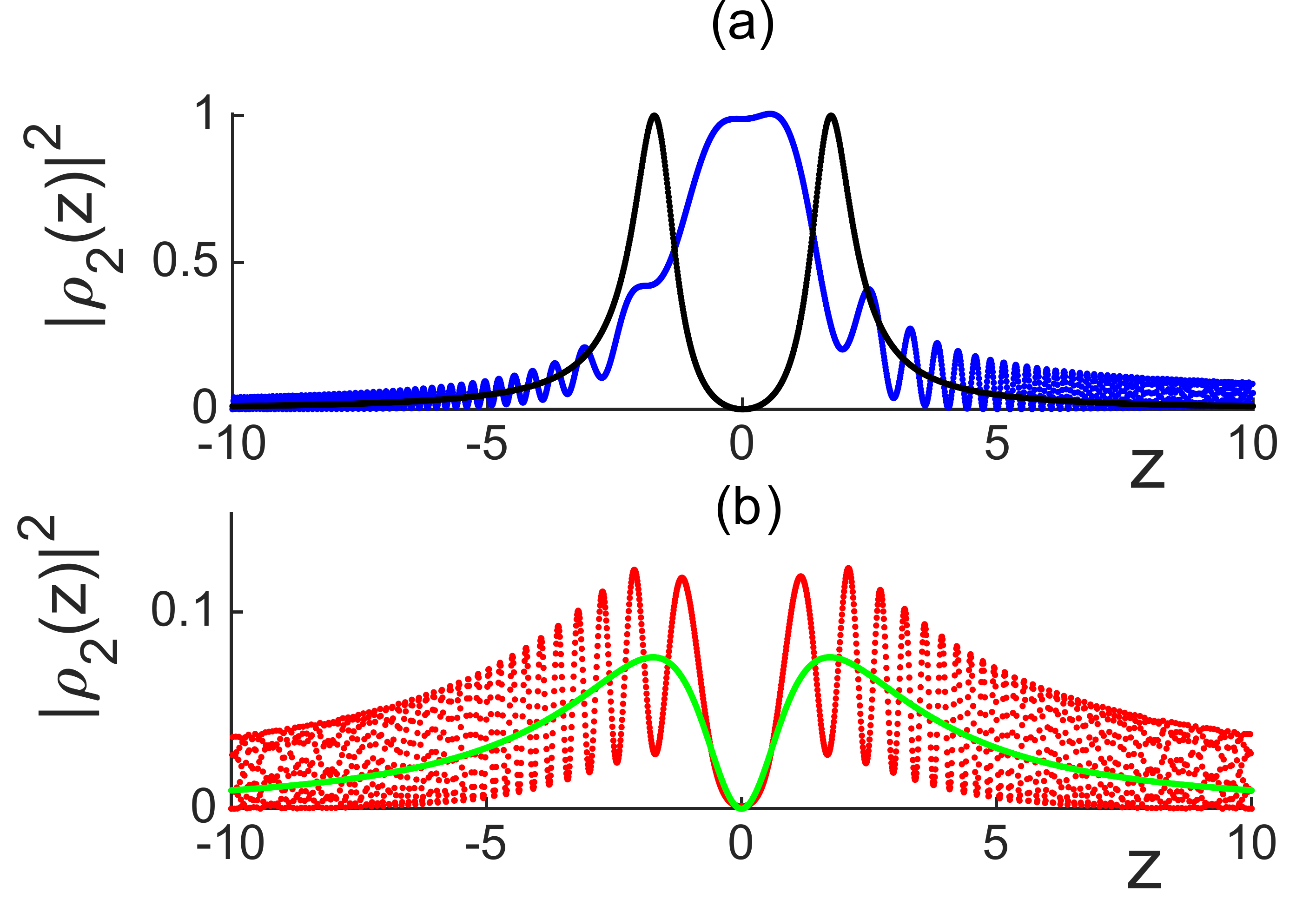}
		\caption{(Color online)    Numerical solution,  $|\rho_2(z)|^2$, of the system   Eq. (\ref {second}) is plotted for two values of
			detuning $\delta=3$ (a) and $\delta =-3$ (b) and for $G=1$. The black line in (a) and the green line in (b) arethe envelops plotted from 
			Eq.  (\ref{differential}). We see that the agreement with theory is much better in (b), since Eq.   (\ref{differential}) neglects interference.  }
		\label{f5}
	\end{figure}

\begin{figure}
	\includegraphics[scale=0.28]{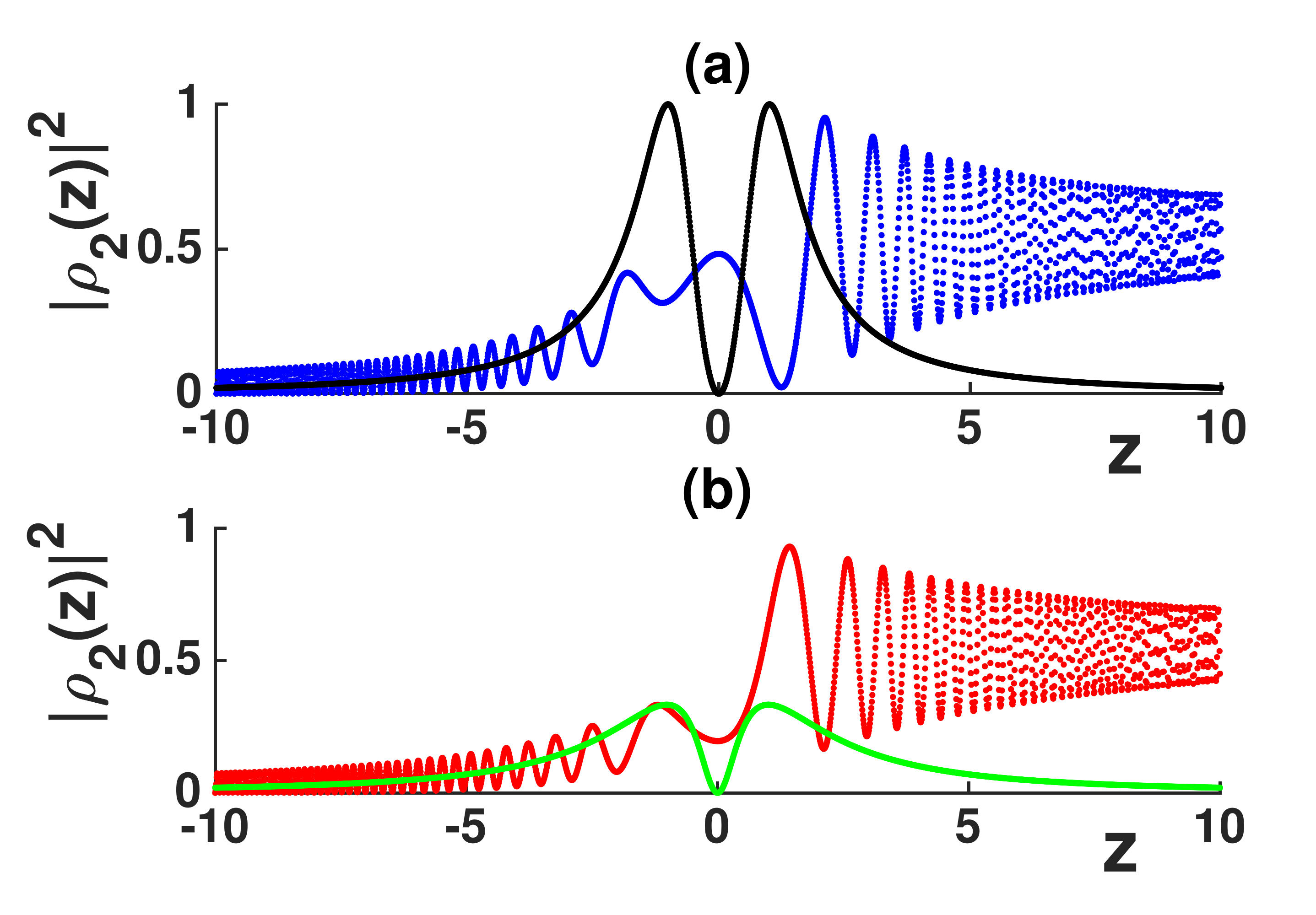}
	\caption{(Color online)    Numerical solution,  $|\rho_2(z)|^2$, of the system   Eq. (\ref {second}) is plotted for two values of
		detuning $\delta=0.5$ (a) and $\delta =-0.5$ (b) and for $G=0.7$. The black line in (a) and the green line in (b) are the envelops plotted from 
		Eq.~ (\ref{differential}).  We see that   $|\rho_2(z)|^2$   approaches finite values at large $z$. This approach
	is accompanied by huge oscillations which complicate the determination of the transmission coefficient.   }
	\label{f5'}
\end{figure}

In Fig. \ref{f5}  we plot $|\rho_2(z)|^2$ for two values of detuning $\delta=3$ and $\delta=-3$ with $G=1$. 
Apparently the smooth part of $\delta=-3$ curve agrees with theoretical prediction      Eq.~(\ref{differential}) much better than
the smooth part of $\delta=-3$ curve. The reason for this is 
obvious:   Eq.   (\ref{differential}) does not capture the interference between the 
virtual transitions at negative and positive $z$. This interference, having the same origin as St{\"u}kelberg 
oscillations takes place at  positive $\delta$.
Since $\delta$ in Fig. \ref{f5}   was chosen to be big, the values of $\rho_2(z)$ approach zero at large $z$.
To capture the finite transmission, we chose the parameters $G=0.7$ and $\delta=\pm 0.5$, 
and plotted $|\rho_2(z)|^2$ in Fig. \ref{f5'}. The agreement with  Eq. (\ref{differential}) is worse in Fig. \ref{f5'}
since, for chosen parameters, the regime of transmission is less ``semiclassical". We also see that approaching
of  $|\rho_2(z)|^2$ to finite values at large $z$ is accompanied by huge oscillations. These oscillations introduce
an uncertainty in the value  $|\rho_2(\infty)|^2$  due to necessary averaging. This uncertainty manifests itself as wiggles
in the dependencies of $|\rho_2(\infty)|^2$ on $G$ and $\delta$ to which we now turn.

\begin{figure}
	\includegraphics[scale=0.32]{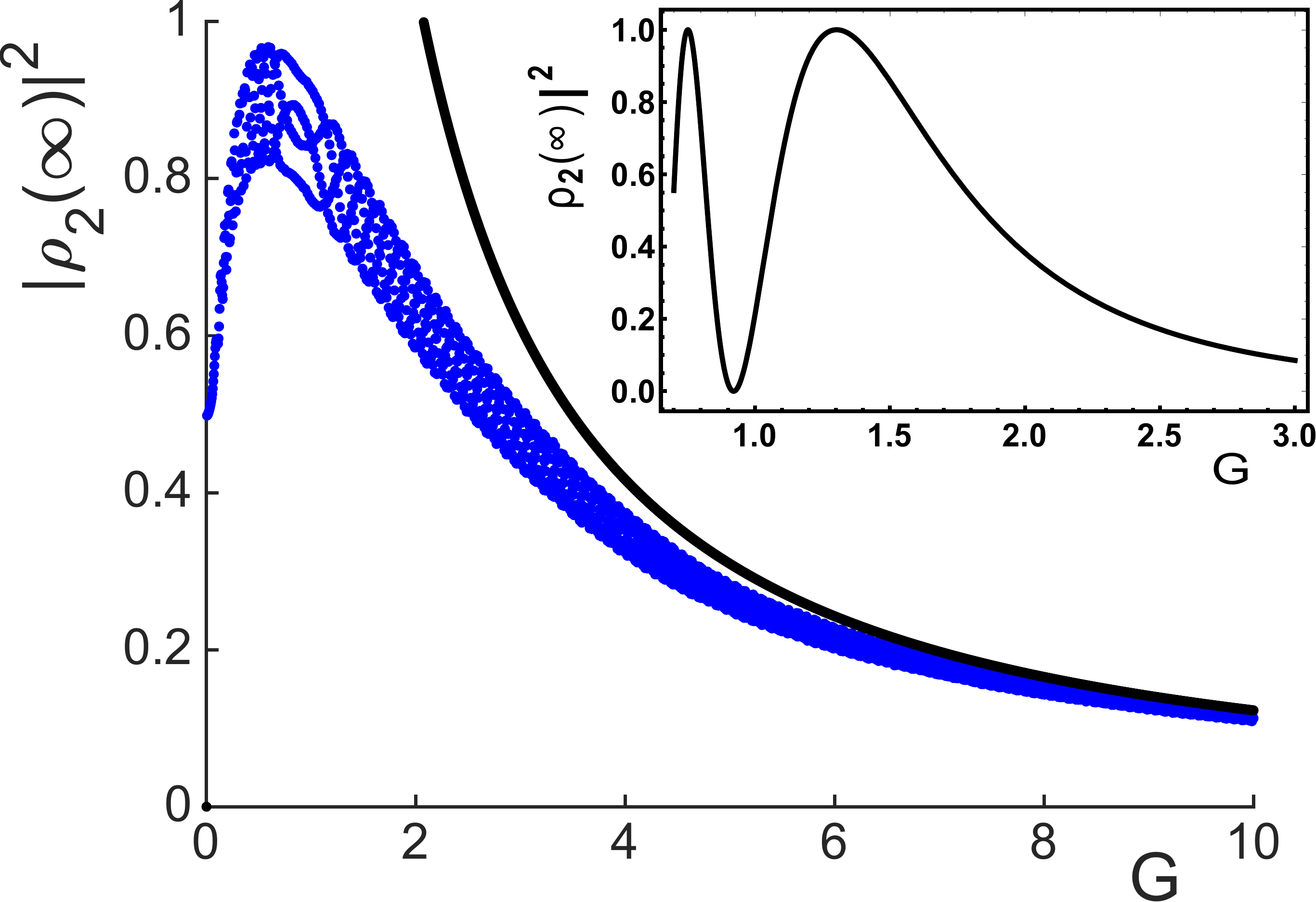}
	\caption{(Color online)    (Blue curve)   Transmision coefficient obtained numerically is plotted versus dimensionless magnetic field, $G$, directly at topological transition $\delta=0$. (Black curve) Theoretical prediction for  transmision coefficient is plotted from Eq. (\ref{Gdependence}). In the inset we plot the prediction,$\sin^2(\frac{8}{3G^2})$, based
	on heuristic argument given in Section V.}
	\label{f6} 
\end{figure}

For zero detuning, the theoretical prediction for the transmission coefficient
is
\begin{equation}
\label{Gdependence}
|\rho(\infty)|^2 =\frac{2.645}{G^{4/3}},
\end{equation}
as follows from Eq. (\ref{Airy1}). In Fig. \ref{f6} we plot this $G$-dependence together with
$|\rho(\infty)|^2(G)$ obtained numerically. We observe the agreement with theory at large $G$, where
the theory is applicable. Concerning the theoretically relevant small-$G$ domain, numerical errors did not
allow us to  establish the $G$-dependence at real small $G$. It can be concluded that the averaged over
strong oscillations transmission coefficient approaches $\frac{1}{2}$  at small $G$ and has a maximum near
$G=1$.  We discuss the theoretical prediction for $|\rho(\infty)|^2(G)$  at small $G$ in the next Section.

%

\begin{figure}
	\includegraphics[scale=0.32]{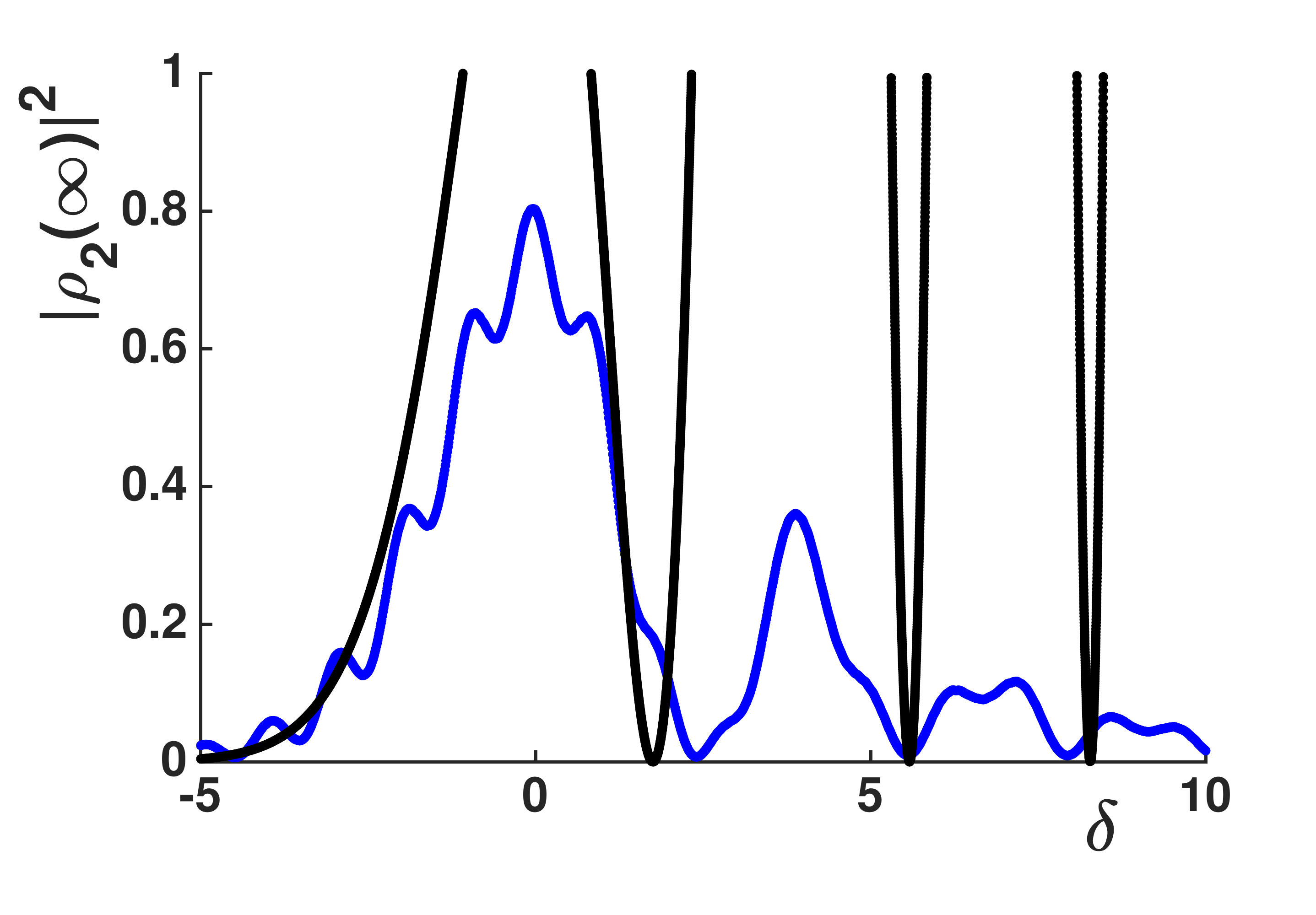}
	\caption{(Color online)   Transmision coefficient (blue curve) obtained numerically is plotted versus the detuning, $\delta$, for dimensionless magnetic field $G=1.5$.  Theoretical prediction for  transmission coefficient
		(black curve) is plotted from Eq.~ (\ref{Airy1}).  Small wiggles in the numerical curves are the artifact of the averaging procedure.  For negative detunings the agreement with theory is good. For positive detunings, theory
		predicts oscillations with magnitude bigger than $1$, while in numerical curve these oscillations, manifesting  the St{\"u}ckelberg interference, slowly decrease with $\delta$.}
	\label{f7}
\end{figure}

Finally, we studied numerically the dependence of the transmision coefficient on detuning, $\delta$. The result is shown in Fig. \ref{f7} for the value of $G=1.5$. We see that for negative $\delta$, the numerics agrees quite well with the theoretical prediction Eq. (\ref{Airy1}). For large positive $\delta$, Eq.~(\ref{Airy1}), strictly speaking, does not apply, but qualitative agreement is apparent. Oscillatory behavior of the transmission coefficient is the consequence of the St{\" u}ckleberg interference of virtual Landau-Zener transitions taking place at $z=\pm\frac{\delta^{1/2}}{G}$.

\section{Discussion}
({\em i}) It is instructive to compare our analysis to Ref. \onlinecite{G3}, where it was concluded that
the probability $|{\cal T}_E|^2$ is given by the Landau-Zener formula. Let us trace how this
formula might emerge from our system Eq. (\ref{system3})

In the semiclassical limit the solutions of the system are ${\tilde \Upsilon_1}$, ${\tilde \Upsilon_2}$ are proportional to $\exp\left[i\sigma(\kappa) \right]$, where the derivative of the action, $\sigma(\kappa)$, is given by
\begin{equation}
\label{action}
\sigma^{\prime}(\kappa)=\pm \frac{{\cal F}_1-{\cal F}_2}{2{\cal F}_1{\cal F}_2}\Bigg[\left(\delta-\kappa^2 \right)^2
+\frac{4\nu^2{\cal F}_1{\cal F}_2}{\left({\cal F}_1-{\cal F}_2\right)^2 }  \kappa^2 \Bigg]^{1/2}.
\end{equation}
Using Eq. (\ref{slopes}),  we specify the combinations  in the square brackets and in the prefactor
\begin{equation}
\frac{4\nu^2{\cal F}_1{\cal F}_2}{\left({\cal F}_1-{\cal F}_2\right)^2 } = (4-\nu^2),~~~\frac{{\cal F}_1-{\cal F}_2}{2{\cal F}_1{\cal F}_2}=\left(\frac{\Delta l}{\alpha}  \right)^2\frac{2\nu}{4-\nu^2}.
\end{equation}

It seems that for small $\delta \ll 1$ the term 
$\kappa^2$ 
can be dropped from $(\delta -\kappa^2)^2$.
Indeed, if this term is dropped, the expression in the square brackets turns to
zero at $\kappa=\pm i\kappa_{-}$, where 
\begin{equation}
\kappa_{-} \approx \frac{\delta}{\left(4-\nu^2\right)^{1/2}}\approx \frac{\delta}{2}.
\end{equation}
Since $\kappa_{-}^2 =\frac{\delta^2}{4}$ is much smaller than $\delta$, dropping $\kappa^2$ is
justified. Once $\delta^2$ is dropped, the expression for $\sigma^{\prime}(\kappa)$ assumes the standard Landau-Zener form with transition probability given by:
$\exp\bigg[-\frac{\pi\nu\delta^2}{8}\big(\frac{\Delta l}{\alpha}   \big)^2    \bigg]$.  This is the result obtained in 
Ref. \onlinecite{G3}.

In our opinion, the flaw of this approach is that, in addition $\kappa=\pm i\kappa_{-}$,
Eq. (\ref{action}) turns to zero at 
$\kappa = \pm i\kappa_{+}$, where 
$\kappa_{+}\approx \left(4-\nu^2\right)^{1/2} \approx 2$. The point $\kappa_{+}$ originates from the second derivatives, $\frac{\partial^2\Phi_{1}}{\partial
	u_1^2}$, $\frac{\partial^2\Phi_{1}}{\partial
	u_1^2}$, 
	in Eq. (\ref{linearized}) which accounts for the curvature of the spectrum neglected in Ref.~\onlinecite{G3}. 
 The value $\kappa_{+} $ is much bigger than $\delta$ and depends on detuning only weakly. 
This suggests that ${\cal T}_E$ is the result of  a ``two-stage" process: one involving big momentum transfer
$\sim \kappa_{+}$ and another involving small momentum transfer, $\sim \kappa_{-}$. The resulting 
${\cal T}_E$ is a strongly oscillating function of detuning and magnetic field. In fact, similar situation, i.e.
numerous complex zeros in $\sigma'$, was encountered in Refs.  \onlinecite{Nakamura1994, Nakamura1995, Montambaux'2012, Montambaux2012,Montambaux2014, Montambaux2015}.

({\em ii} )
Overall, we were not able to capture the most relevant domain where both $\delta$ and $G$  are small neither
analytically nor numerically. This is due to strongly oscillating character of $|\rho_2(z)|^2$.   The physical
origin of this complication is that simple linearizng the Fermi contours near the crossing is insufficient for
finding the transition probability.  The amplitudes $\rho_1(z)$ and $\rho_2(z)$ keep ``talking" to each other
outside the domain where linearization applies.
Below we present a  heuristic account of  the behavior of the  transmission coefficient  at zero detuning.
Conventionally,\cite{Dykhne}  the transmission coefficient in the Landau-Zener problem can be found 
upon setting $\kappa$  in the expression for $\sigma'(\kappa)$ to be  purely imaginary and integrating
between two turning points. This procedure is applicable when the resulting action is big, so that the 
transmission is small. If we adopt this procedure in Eq. (\ref{action}) after setting $\delta=0$, we would 
realize that, unlike Landau-Zener transition the action is {\em imaginary} and is equal to  
 \begin{equation}
 \label{Sinoscillation}
i\sigma =\int\limits_{-2/G}^{2/G}dz\left[z^2-\frac{G^2z^4}{4}    \right]^{1/2}=\frac{8}{3G^2}.
 \end{equation}
We see that at small $G$ the magnitude of action is big and that the transmission coefficient  {\em oscillates} with
$G$ instead of being exponentially small. We cannot judge about the prefactor, except that in Landau-Zener
transition the prefactor is $1$. This leads to the prediction $|\rho_2(\infty)|^2=\sin^2(\frac{8}{3G^2})$. This prediction is plotted in the inset of Fig.~\ref{f6}. A maximum at $G=1.3$ can possibly account for the behavior
of the numerical curve around $G\sim 1$. If the above heuristic argument applies, then the $\delta$-dependence of the transmission at small $G$ should be weak.

({\em iii}) The result Eq. (\ref{Airy})
can be derived directly from the system Eq. (\ref{linearized}) without transforming to
the momentum space.  The zeroth-order solution 
 of the first equation is
 $Ai\left[ \frac{1}{{\cal F}_1^{1/3}}\left( {\delta}{{\cal F}_1}-u_1\right)\right]$. 
Thus, the right-hand side in the second equation
is the derivative of the Airy function. Forced
solution of the second equation contains the
overlap of this right-hand side with the free solution
of the second equation, which is $\Phi_2(u_1)= Ai\left[ \frac{1}{{\cal F}_2^{1/3}}\left( {\delta}{{\cal F}_2}
-u_1\right)\right]$.
Then the  result  Eq. (\ref{Airy}) follows from the identity
\begin{eqnarray}
\label{identity}
\int\limits_{-\infty}^{\infty}dx Ai\left[ \lambda(x-a)\right] 
Ai^{\prime}\left[ \mu(x-b) \right]\nonumber\\
=\frac{\pi\lambda}{\left(\lambda^3-\mu^3  \right)^{2/3}}Ai^{\prime}\left[\frac{\lambda\mu(a-b)}{\left(\lambda^3-\mu^3   \right)^{1/3}}   \right],
\end{eqnarray}
which can be easily verified using the integral representation of the Airy function.

\section{Acknowledgements}
This work was supported by the Department ofEnergy, Office of Basic Energy Sciences, Grant No. DE-FG02-06ER46313.


\end{document}